\documentclass[journal,final,11pt,journal]{IEEEtran} 
\usepackage[T1]{fontenc}

\usepackage{balance}
\usepackage{xcolor}
\usepackage{soul}
\usepackage[caption=false]{subfig}
\usepackage{algorithm}
\usepackage{algorithmic}
\usepackage{multicol}
\usepackage{cite} 
\usepackage{amsmath}

\usepackage[cmintegrals]{newtxmath}
\usepackage{multirow}
\colorlet{shadecolor}{yellow}
\usepackage{array}
\newcolumntype{P}[1]{>{\centering\arraybackslash}m{#1}}
\definecolor{Gray}{gray}{0.85}
\newcolumntype{a}{>{\columncolor{Gray}}c}
\usepackage{color}

\usepackage[pdftex]{graphicx}
 
\usepackage{glossaries,color}

\newacronym{3GPP} {3GPP} {third generation partnership project}
\newacronym{embb}    {eMBB}    {enhanced mobile broadband}
\newacronym{mmtc} {mMTC} {massive machine type communication}
\newacronym{urllc} {URLLC} {ultra reliable and low latency communication}
\newacronym{ran} {RAN} {radio access network}
\newacronym{cran} {C-RAN} {cloud - radio access network}
\newacronym{rrm} {RRM} {radio resource management}
\newacronym{sla} {SLA} {service level agreement}
\newacronym{multirat}{Multi-RAT}{multiple radio access technologies}
\newacronym{snssai} {S-NSSAI} {single-network slice selection assistance information}
\newacronym{oam} {OAM} {operations administration and maintenance}
\newacronym{sd} {SD} {slice differentiator}
\newacronym{fwa} {FWA} {fixed wireless access}
\newacronym{rg} {RG} {residential gateway}
\newacronym{mec}{MEC} {multi-access edge computing}
\newacronym{qos}{QoS} {quality of service}
\newacronym{sudep}{SUDEP} {sudden unexpected death in epilepsy}
\newacronym{sdn}{SDN} {software-defined networking}
\newacronym{nfv}{NFV} {network-function virtualization}
\newacronym{wlan}{WLAN} {wireless local-area network}
\newacronym{ofdma}{OFDMA} {Orthogonal Frequency Division Multiple Access}
\newacronym{sta}{STA} {station}
\newacronym{ap}{AP} {access point}
\newacronym{gtc}{GTC}{generalized tonic-clonic}
\newacronym{ue} {UE} {user equipment}
\newacronym{ai} {AI} {artificial intelligence}
\newacronym{iot}{IoT}{Internet-of-things}
\newacronym{wban}{WBAN}{wireless body area network}
\newacronym{gcs}{GCS}{gateway computing server}
\newacronym{lte}{LTE}{long-term evolution}
\newacronym{ble}{BLE}{Bluetooth low energy}
\newacronym{eeg}{EEG}{electroencephalograph}
\newacronym{e2e}{E2E}{end-to-end}
\newacronym{m-lwdf}{M-LWDF}{modified-largest weighted delay first}

\usepackage[absolute]{textpos}
\usepackage{everyshi}

\begin{document}

\author{Sergio Martiradonna, Giulia Cisotto, Gennaro Boggia, \\ Giuseppe Piro, Lorenzo Vangelista, and Stefano Tomasin

\thanks{ S. Martiradonna, and G. Piro are with the Department of Electrical and Information Engineering, Polytechnic of Bari, Italy. G. Cisotto, L. Vangelista, and S. Tomasin are with the Department of Information Engineering, University of Padova, Italy. G. Cisotto is also with the Integrative Brain Imaging Center, National Center of Neurology and Psychiatry, Tokyo, Japan.}
}

\title{Cascaded WLAN-FWA Networking and  Computing Architecture for Pervasive In-Home Healthcare}

\maketitle

\begin{abstract}
Pervasive healthcare is a promising assisted-living solution for chronic patients. However, current cutting-edge communication technologies are not able to strictly meet the requirements of these applications, especially in the case of life-threatening events. To bridge this gap, this paper proposes a new architecture to support indoor healthcare monitoring, with a focus on epileptic patients. Several novel elements are introduced. The first element is the cascading of a WLAN and a cellular network, where IEEE 802.11ax is used for the wireless local area network to collect physiological and environmental data in-home and 5G-enabled Fixed Wireless Access links transfer them to a remote hospital. The second element is the extension of the network slicing concept to the WLAN, and the introduction of two new slice types to support both regular monitoring and emergency handling. Moreover, the inclusion of local computing capabilities at the WLAN router, together with a mobile edge computing resource, represents a further architectural enhancement. Local computation is required to trigger not only health-related alarms, but also the network slicing change in case of emergency: in fact, proper radio resource scheduling is necessary for the cascaded networks to handle healthcare traffic together with other promiscuous everyday communication services. Numerical results demonstrate the effectiveness of the proposed approach while highlighting the performance gain achieved with respect to baseline solutions.

\end{abstract}

\begin{textblock*}{17cm}(1.7cm, 0.5cm)
\noindent\scriptsize This work has been submitted to the IEEE for possible publication. Copyright may be transferred without notice, after which this version may no longer be accessible.
\textbf{Copyright Notice}: \textcopyright 2020 IEEE. Personal use of this material is permitted. Permission from IEEE must be obtained for all other uses, in any current or future media, including reprinting/republishing this material for advertising or promotional purposes, creating new collective works, for resale or redistribution to servers or lists, or reuse of any copyrighted component of this work in other works.
\end{textblock*}

\IEEEpeerreviewmaketitle

\sloppy
\section*{Introduction}
\emph{Pervasive healthcare} envisages a continuous and ubiquitous monitoring of physiological signals and vital parameters, while improving the living conditions of patients at their homes. Especially for patients in the most critical conditions, e.g., patients suffering from severe epilepsy, continuous monitoring is crucial for effective life-saving interventions in case of emergency, and to prolong life expectancy. 
For example, up to a third of all premature deaths worldwide are either directly or indirectly attributed to epilepsy~\cite{VanDerLende2016}.
Particularly, \gls{sudep} is a direct cause of death occurring for $1-2$ every $1\,000$ severe epileptic patients per year, and it is estimated to occur in one  every $2\,000-5\,000$ \gls{gtc} seizures, a particular kind of severe convulsive seizure~\cite{VanDerLende2016}.
It has already been shown that continuous daily and overnight monitoring of patients can reduce the frequency of seizures and provide immediate protective mechanisms, thus reducing the risk of \gls{sudep}. The state of the art in clinical monitoring of epileptic patients is performed by specialized caregivers, with the help of a $3$D video camera and an \gls{eeg} that measures brain activity. However, the cost of such continuous supervised monitoring has been estimated in the order of thousands of dollars per seizure~\cite{VanDerLende2016}.
Therefore, autonomous decision-support systems for epilepsy management in smart homes represent promising assisted-living solutions for the near future.
Thus, in this work, we target a healthcare solution for both continuous monitoring and emergency handling in severe epilepsy. 
However, the proposed system could be applied also to patients sharing a similar need for pervasive healthcare, e.g., affected by cardiovascular or chronic diseases, or doing rehabilitation at home.

Current cutting-edge communication technologies, including IEEE 802.11ax  \gls{wlan}, fifth-generation (5G) mobile networks, and its \gls{fwa} solution offer a broadband backhaul connectivity even in suburban areas and can support pervasive healthcare at large scale. Unfortunately, baseline implementations do not provide any priority to healthcare messages over other applications in case of life-threatening events. The emerging \emph{network slicing} paradigm, as applied to 5G \cite{TS22261}, to \gls{wlan}\cite{richart2019slicing}, and to heterogeneous architectures exploiting \gls{wlan} technology at the radio interface, and to 5G components in the core network \cite{carmo2019network}, fulfills \gls{qos} constraints through traffic isolation (e.g., by assigning virtual resources under a common communication infrastructure).
This approach, however, has been scarcely investigated in the healthcare context so far,
especially to tackle life-threatening events.


In this paper we propose a new network solution for indoor healthcare monitoring, in particular for epileptic patients. The architecture is composed of various new elements. First, we cascade a WLAN and a cellular network, where IEEE 802.11ax is used in-home and 5G-enabled Fixed Wireless Access links transfer them to a remote hospital: this solution is flexible and particularly suitable to serve remote areas, where a fiber link is not available. Second, in order to support both regular monitoring and emergency handling, we introduce two new slice types and extend the cellular network slicing concept to WLAN. Third, we propose to use an enhanced WLAN router with local computing capabilities, which is still controlled by the cellular network. The latter, integrated with mobile edge computing resources, makes the resulting architecture more flexible and powerful. Indeed, the local computation capabilities can be exploited to trigger health-related alarms and dynamic network slicing in case of emergency, and to provide resource scheduling of both healthcare traffic and  other promiscuous everyday communication services. Lastly, we demonstrate the performance of the resulting architecture and compare them to baseline solutions.

\section*{Epilepsy Management}~\label{sec:ehealthscenario}

\begin{table*} 
	\caption{Communication Requirements of Monitoring Devices} \label{tab:scenarios}
	\centering{
		\small
		\begin{tabular}{ | P{2.5cm} || P{2.5cm}| P{1.7cm} | P{1.0cm}| P{2cm} | P{2.0cm} |}
			   \hline						
			
			\hline 
			\hline 
			\textbf{Data type} & \textbf{Sub-type (no. channels)} & \textbf{End-to-end latency} & \textbf{Jitter} & \textbf{Survival Time} & \textbf{Data rate (aggregated)} \\
			
			&     & ms & ms & ms &   \\
			
			\hline 
 			\multicolumn{6}{c}{\textbf{Regular Monitoring (standard video EEG)}}\\			
 			
			\hline
			
			Multimedia & 3D camera 1 & $150$ & $30$ & $180$ & $10$~Mbps  \\  \hline						
			
			Electrophysiology & EEG (30) & $250$ & $25$ & $175$ & $1$~Mbps  \\    \hline						
			
			\hline 
 			\multicolumn{6}{c}{\textbf{Emergency Monitoring (additional data)}}\\

			\hline
			
			Multimedia & 3D camera  2 & $150$ & $30$ & $180$ & $10$~Mbps \\      \hline
			
			
			& Speaker & $150$ & $25$ & $175$ & $220$~kbps  \\        \hline
			
			Electrophysiology & ECG (3) & $250$ & $25$ & $275$ & $0.5$~Mbps  \\    \hline
			
			
			& EMG (4) & $250$ & $25$ & $275$ & $0.5$~Mbps \\    \hline
			
			Optics            & SpO$_2$ & $250$ & $25$ & $275$ & $0.5$~Mbps  \\    \hline

			Vitals & Temperature & $250$ & $25$ & $275$ & $100$~kbps   \\         \hline
			
			& Blood pressure & $250$ & $25$ & $275$ & $100$~kbps  \\      \hline
			
			& Heart rate & $250$ & $25$ & $275$ & $100$~kbps   \\          \hline
			
			& Respiration rate & $250$ & $25$ & $275$ & $100$~kbps  \\    \hline\hline
			
		\end{tabular}
	}
	~\\[0.05cm]
	\label{tabella_2}
\end{table*}



Recently, other architectures have been presented for continuously managing patients in critical conditions, e.g., severe epileptic patients, at their homes. Most of them employ wearables and portable devices to collect vitals and brain signals, as well as context information, i.e., the patient's location. The most common solution includes also a \gls{mec} server for data analysis using \gls{ai} algorithms: e.g., in~\cite{Hosseini2020}, the authors propose an architecture based on \gls{lte} and \gls{sdn}, where an edge gateway is assisted by \gls{ai} in the localization of the epileptic foci in the brain, and delivers effective real-time brain stimulus regulating the epileptic activity while mitigating symptoms. Deep learning algorithms running on a \gls{mec} server have been advocated to support the early prediction of epileptic seizures. Although promising, these solutions still lack a realistic in-field deployment (e.g., multiple users, longer distances).

In~\cite{Vergara2017} a two-hop monitoring architecture is proposed, which collects 3D accelerometer traces and the heart rate through a smart bracelet. The latter sends data via \gls{ble} to a smartphone, which acts as local gateway to Internet via \gls{wlan}. The minimum end-to-end latency is $175$~ms when serving a single user.

In~\cite{Sareen2016}, the authors propose a cloud-based seizure prediction architecture including a \gls{wban}, a GPS-based localization  and a localization software hosted in an Amazon elastic compute cloud instance. Several \gls{ai}-based algorithms have been tested in this study to detect and predict seizures from a low-cost wireless \gls{eeg} headset. However, the architecture covers a short-range area that can not be considered as a realistic scenario.  In~\cite{Asif2018}, edge computing is also proposed to deliver real-time alarms and improve user interaction during emergency in other \gls{iot}-healthcare scenarios, e.g., for preventing falls of elderly people and in mobile healthcare units. 

Interestingly, the \emph{HealthEdge} project~\cite{HealthEdge2017} suggests to prioritize two different types of traffic, i.e., \emph{human behaviour} and \emph{health emergency}, and to decide whether to pre-process data in a \gls{mec} server or to send them directly to the cloud. Task scheduling in~\cite{HealthEdge2017} has been implemented at the edge workstation with benefits on both the bandwidth utilization and the total task processing time. In that case, the edge scheduler directs traffic to  either the \gls{mec} or the cloud, solely based on the patient's physiological data and no other context information, e.g., current network traffic, is taken into account. At the same time, no strategy for dynamic switching between the two types of traffic has been investigated. 

To the authors' knowledge, the literature has not yet investigated the use of specific e-health slice types, and solutions for epilepsy management (or similar scenarios as considered in this paper) are not available. Still, as network slicing is emerging as a new solution, resource management, admission control, and traffic prioritization in the \gls{ran} are relevant problems \cite{9003208}.

\subsection*{Requirements Definition}

\begin{figure}[!t]
	\centering
    \includegraphics[width=1\hsize]{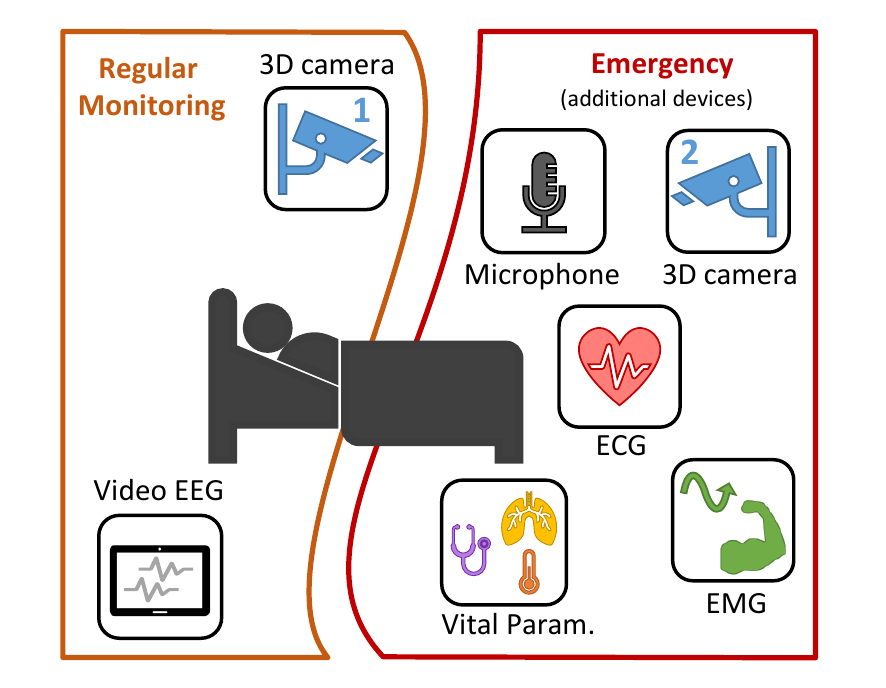}
	\caption{Data collection setup in case of severe epilepsy management, both during regular monitoring and emergency handling.}
	\label{fig:usecase}
\end{figure}

The case targeted in this paper encompasses, at the same time, two healthcare services: a regular monitoring with mild communication requirements in terms of latency, packet drop, and data rate, and an emergency handling with strict requirements, activated during life-threatening events. Fig.~\ref{fig:usecase} shows the data collection setup of this particular use case.


In \emph{regular monitoring}, we consider the state-of-art \gls{eeg}-video acquisition setup~\cite{VanDerLende2016}, with recording from a $3$D camera and $30$ \gls{eeg} channels. 
During an \emph{emergency}, we assume that an alarm has been triggered based on abnormal \gls{eeg}-video data, and we extend the interaction with the patient by adding a 3D camera, a speaker, 3-leads ECG, 2 bipolar electromyography (EMG) channels, a pulse oximeter to measure peripheral oxygen saturation (SpO$_2$), and a system to acquire the most important vital parameters. This provides the remote specialized clinicians with a better understanding of the situation which, in turn,  highly improves the accuracy in detecting \gls{sudep} for a more effective and early intervention.

Table~\ref{tab:scenarios} summarizes the service requirements, taking into account \gls{qos} metrics typically used in the design of 5G systems \cite{TS22261, Cisotto2020}. 
It is important to highlight that the values reported in Table I refer to the communication delays expected during the run-time phase of a network slice instance.
Moreover, the additional latency related to the activation and configuration of a new slice is experienced only once. Also, the latter is not correlated to the survival time in the run-time phase (see Table~\ref{tab:scenarios}) and it is expected to be much smaller than it.

\section*{Reference Architecture and \\ Proposed Slice Management}~\label{sec:dynslicing_health}
\begin{figure}[!t]
	\centering
    \includegraphics[width=1\hsize]{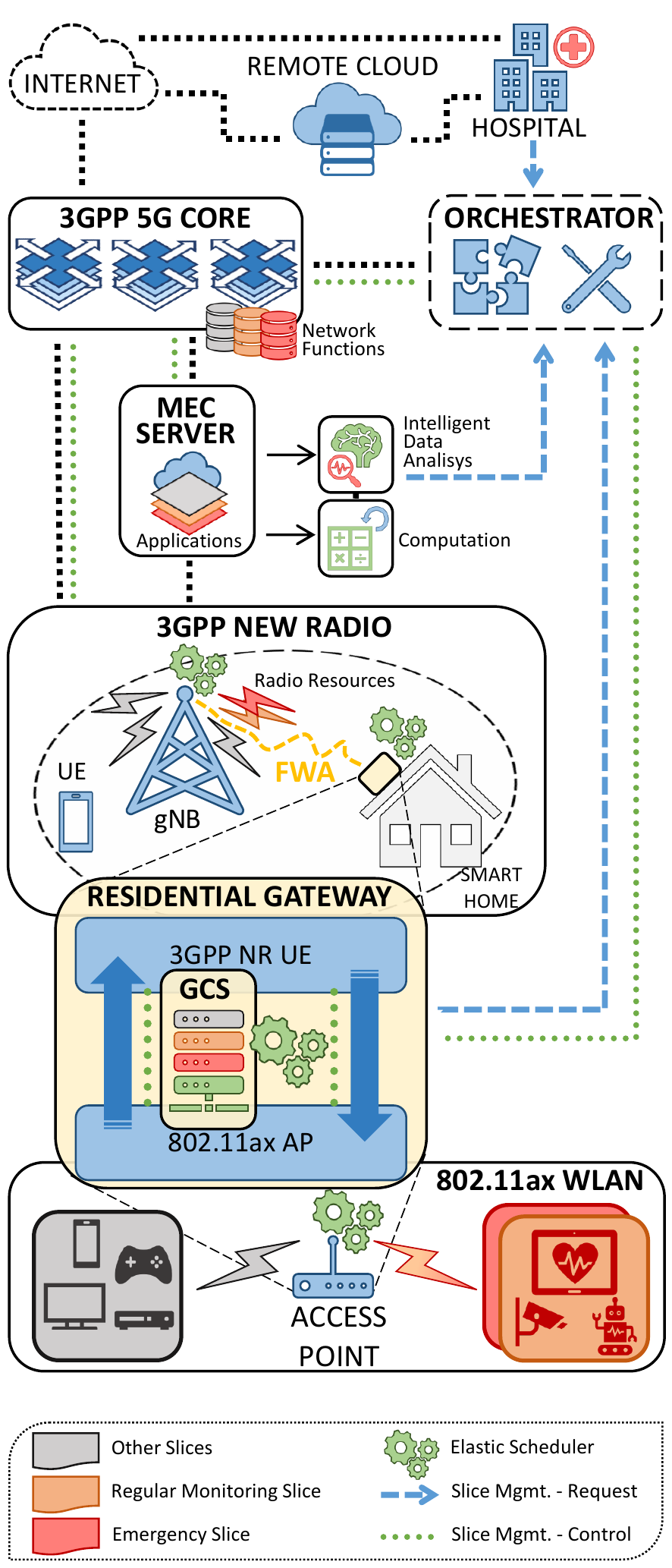}
	\caption{High-level overview of the proposed architecture. }
	\label{fig:architecture}
\end{figure}

In our scenario, patients' smart homes are covered by an IEEE 802.11ax \gls{wlan} connected to the Internet by an \gls{fwa} over a 5G cellular network. As a matter of fact, 5G \gls{fwa} is the most promising alternative, from the techno-economic point of view, to close the digital divide, for urban and suburban areas as well. For the healthcare purposes, IEEE 802.11ax devices collect physiological and environmental data in the patient's home and 5G-enabled \gls{fwa} links transfer them to the remote hospital's cloud servers.

Fig.~\ref{fig:architecture} depicts the proposed connectivity architecture.

As both \gls{wlan} and \gls{fwa} networks are also used for other communication purposes in the smart home, no dedicated network deployment is necessary. Still, in order to protect the new services, the latter are mapped into two new slice types: a \emph{regular monitoring slice-type} when the patient's conditions are stable and an \emph{emergency slice type} for communications during the emergency. The two slices are defined for the cascaded \gls{wlan}-\gls{fwa} networks: therefore, the \gls{wlan} router must support the in-home network slicing and should be equipped with a \gls{gcs}.

Since network slice types are different in their nature (in terms of targeted QoS levels, traffic, network functions, number of nodes involved, etc.) and in their use of a subset of resources available at the radio interface, it is possible to tailor, for each of them, a customized radio resource management scheme that meets their specific requirements. The architecture proposed in our work fully exploits this key capability and natively assumes to implement customized radio resource management schemes for each network slice type.

\subsection*{Cellular Network}

We consider a 3GPP 5G system including 5G NR, the 5G Core network (5GC), and an orchestrator entity.
5G NR is the radio access network, which provides connectivity to each \gls{ue} by base stations, i.e., gNBs, while  5GC creates Packet Data Unit sessions between \glspl{ue} and the Internet.
The orchestrator handles the configuration of the 5G network, e.g., it controls the slice life cycle, instantiates slice resources, and assigns traffic routing policy and flow priorities.  
The \gls{mec} server at the edge of the network  provides the required computational and storage resources to implement functions for real-time performance, as well as to virtualize specific applications.
Specifically, it implements advanced applications for monitoring, classifying, and predicting patients’ behaviors, and it supports to the orchestrator in the real-time management and configuration of slices and resources.

Among \glspl{ue}, there are some \glspl{rg} providing broadband connectivity to smart homes through  FWA links.
In particular, an \gls{rg} acts as a gateway between each wireless connected device in the house and the external 5G network. In other words, the \gls{rg} includes a 3GPP 5G \gls{ue} and an IEEE~802.11 \gls{ap}. Besides, our architecture includes a novel \gls{gcs} in the \gls{rg}, in order to both offload computing task for the \glspl{sta} and manage the interaction at the 5G-802.11 interface. The \gls{gcs} plays the role of \gls{mec} at an even more local level, providing computational capabilities even closer to the end-user applications, still remaining under the full control of the operator and the service provider. Being at the border between the \gls{wlan} and the  cellular network, it both responds quicker to in-home events and has a privileged access to  both \gls{wlan} and cellular network resources, including  \gls{mec}, e.g., for   computations on larger databases. \gls{mec} and \gls{gcs} can also share the workload needed to detect anomalies in the \gls{eeg} monitoring, and to trigger the activation of the emergency slice. 
\subsection*{Wireless Local Area Network}

We choose IEEE~802.11ax as the \gls{wlan} standard since, among other interesting features, it provides centralized scheduling \cite{standard80211ax}. 
Furthermore, we aim at extending the concept of network slicing also in the IEEE~802.11ax network. Indeed, the proposed architecture enables the creation and the dynamic control of network slices spanning from the 5G network to  \glspl{wlan}.
A slice-oriented approach provides isolated and independent resources (e.g., radio resources in both \gls{wlan} and 5G radio networks, computing resources at the edge, dedicated network functions in 5GC) to each network slice. This significantly extends the possibilities of legacy cellular technologies for realizing \gls{qos} differentiation. 
On one hand, as previously mentioned, each network slice type can benefit from a tailored management scheme of its resources, which may efficiently increase the performance with respect to a one-size-fits-all mechanism.
On the other hand, applications requiring \gls{ai}-based heavy computations with stringent time constraints, e.g., advanced healthcare services, can be offloaded to \gls{gcs} and \gls{mec}.

In addition to the slice management in the cascaded networks, we adopt an \emph{elastic} radio resource scheduling which operates on both radio technologies, in a distributed and coordinated manner.
 
\section*{Network Slicing Solution}

Under stable patient's conditions, data collected by the biometric sensors inside the smart home reach the nearest \gls{mec} server through the \emph{regular monitoring} slice. The patient's health-related data are analyzed and processed in the \gls{mec} server. When an emergency occurs, the \gls{rg} should be enabled to transmit and receive data on the \emph{emergency} slice. In this case, not only {\em emergency} data is massively collected to better formulate a diagnosis, but also {\em control} data could be sent from the \gls{mec} to the patient's home, e.g., to alert local caregivers.

\subsection*{Dynamic Slicing}

As the \emph{emergency} slice type, i.e., a high-priority slice type, is rarely instantiated, it may be inefficient to leave it active, thus penalizing the lower-priority traffic and wasting computing resources on the \gls{mec} server. Indeed, while 5G promises to further increase data rates and use new parts of the spectrum, a tremendously higher number of devices is expected to join the network, with a total high band requirement, although with different timing constraints. As a consequence,  resources will be actually constrained and a cautious use of resources is needed. We consider instead a {\em dynamic slicing} approach, wherein a slice can change its type in correspondence of certain events. In this context, a \emph{regular monitoring} slice type is instantiated to carry physiological patient's data and environmental measurements. However, upon a relevant event, e.g., a significant change of any vital parameter or any other life-critical event, the existing slice is promoted to the \emph{emergency} type. This change can be triggered either by the device or by the network. In this latter case, for instance, a certain slice's application, which is virtualized on \gls{mec}, may detect an anomaly in regularly monitored vital parameters exploiting \gls{ai} algorithms.
Then, when critical events occur, the \gls{mec} server should reach the orchestrator to force the change of slice type to {\em emergency}. Indeed, for resource management, we still rely on the orchestrator of the 5G network (shown also in Fig.~\ref{fig:architecture}), as from the standard. 
Note that letting the network change the slice type allows for the use of legacy equipment not directly capable of requesting the setup of new slices, as the slice type is set by the network. Interestingly, {\em dynamic slicing} provides efficient management not only for slice activation, but also for its deactivation, when emergency is solved. Finally, we highlight that the existing literature on dynamic network slicing considers the dynamic creation of new slices as limited to the current 3GPP paradigm, i.e., for slowly reactive systems (see for example \cite{8627115}).


\subsection*{Elastic Resource Scheduling}
In order to meet the \gls{qos} requirements on an end-to-end basis, we envision   that the \gls{rg} controls the scheduling of the \gls{wlan}, thus enforcing the slice requirements inside the smart home, too. 
As a matter of fact, the \gls{wlan} component is typically unaware of the 5G network slicing. Indeed, both \gls{ap} and \glspl{sta}' knowledge of the network is limited to the \gls{wlan}, hence ignoring how different slices are configured in the 5G segment. 
It is thus necessary that the \gls{gcs} implements a mapping function that translates the requirements of a specific slice/service type into a \gls{wlan} service class and efficiently schedules and manages the traffic flows within the smart home. For instance, packets of the emergency slice are assigned a higher priority, in order to ensure reliable and low-latency services in critical situations. In this way, the network slicing should be enforced for the cascaded IEEE~802.11ax and 5G networks.

 
The slice \gls{qos} requirements are satisfied by a novel \emph{elastic} radio resource scheduling policy.
Thanks to the \gls{gcs}, the radio resource management policy handles different queues within the \gls{rg}, one for each slice.
Moreover, the radio resource allocation must satisfy the requirements on both the maximum end-to-end latency and the survival time (see Table~\ref{tabella_2}).
For this reason, within both the IEEE~802.11ax and 5G networks, the scheduling algorithm distributes the radio resources taking also into account the queuing delay across the cascaded network.
Therefore, our solution requires the devices to communicate the experienced queuing delay, along with their buffer status report.

In uplink, as soon as a packet is correctly received by the \gls{ap} and the \gls{gcs} pushes it in the related queue of the \gls{rg}, its accumulated delay is tracked. Each queue is sorted according to the packet delay and expired queued packets are dropped. As a result, the accumulated delay in the two-tier segment is taken into account in scheduling resources.

Furthermore, the envisioned scheduler  determines the number of radio resources requested by each flow, hence by each slice, in a given time window $T$. This evaluation is based on the agreed QoS parameters (e.g., average transmission rate), as well as on the channel conditions experienced by the users, and it is conducted in both networks, i.e., in the radio access network and in the \gls{wlan}.
When the amount of requested radio resources can be satisfied  in $T$, the scheduling follows the \gls{m-lwdf} approach. Since the delay of each packet is tracked by the \glspl{rg}, the \gls{m-lwdf} scheduler aims at satisfying the \gls{qos} requirements on an end-to-end basis. 
Conversely, when the number of available radio resources in $T$ does not  match the requests, an elastic resource scaling is first applied. The elastic scaling proportionally reduces the number of available radio resources for each active slice instance in $T$, according to the resources surplus requested by each slice.
After scaling, slices are  grouped into two sets, i.e., with priority and without priority. Then, in each scheduling interval, radio resources are assigned first to packets of slices with priority, and then to those with no priority.

\section*{A Case Study}~\label{sec:results}
 
We model a European suburb according to the reference \gls{fwa} scenario of \cite{ericsson2019fwa}, including 1\,000 households per km$^2$ and a grid of three-sector macro sites with an average inter-site distance of about 1~km. The network is designed to connect simultaneously up to 30\% of the covered households, thus each sector serves approximately 88 households. Each sector is equipped with 64 transmit/receive antennas, working in the sub-6~GHz band and each \gls{rg} has 2 transmit and 4 receive antennas. Therefore, up to 16 different spatial layers may be multiplexed on a single resource block. All households also generate uplink traffic for a generic \gls{embb} slice according to the models in \cite{802sim}.

As about $1.8$ million people in Europe with epilepsy is at risk of SUDEP, there are 2 patients per sector, on average, to manage. 
In the worst-case scenario, we assume one epileptic patient at high risk of SUDEP and another epileptic patient with no risk of SUDEP to be monitored at the very same time in a single sector. Healthcare traffic flows are generated according to the specifics of Table~\ref{tab:scenarios}.


In each household, we consider single-user MIMO (SU-MIMO) for IEEE~802.11ax, with a single spatial stream per STA. We do not model legacy STAs in the network, although we assume that the \gls{ap} reserves 30\% of the time for legacy transmissions and extra signal processing delay. 
Based on the channel conditions, the APs and the \gls{rg} select the appropriate modulation and coding schemes, in order to guarantee a target maximum block error rate (BLER) through link adaptation. Moreover, both the transport block size for 5G 
and the data rate for 802.11ax networks 
are set accordingly, following the standards.
We consider a higher BLER value for the IEEE~802.11ax link, in order to take into account the interference and possibly busy channels, as each STA performs carrier sensing and the transmission is canceled whenever the medium is busy, resulting in transmission delays.


The proposed scheduling solution (reported in the following as \emph{Elastic}) has been compared with: 
\begin{itemize}
    \item a solution performing no slicing at all (reported in the following as \emph{Basic}), without any proper resource scheduling policy, i.e., proportional fair scheduling is used at all nodes;
    \item a network slicing solution (reported in the following as \emph{\gls{e2e}}), including slicing within the \gls{wlan} and the use of the \gls{m-lwdf} scheduler for each slice, although without the  elastic resource scheduling, i.e., neither traffic prioritization, expired packets management, nor resource scaling.
\end{itemize}  

By comparing these solutions, it is possible to appreciate the advantages of using network slicing and the introduced elastic resource scheduling.

The impact of scheduling strategies over the end-to-end latency, the communication service availability, and the number of emergency slice packets meeting the required \gls{qos},  has been investigated by computer simulations. In the following, $\alpha$ and $\beta$ are weights to further manage priorities associated with the \gls{embb} slice and the healthcare slices, respectively: a higher value of these parameters means a higher priority of the corresponding slice.

\subsection*{End-to-End Latency}

\begin{figure}
 \centering
 \includegraphics[width=\linewidth]{{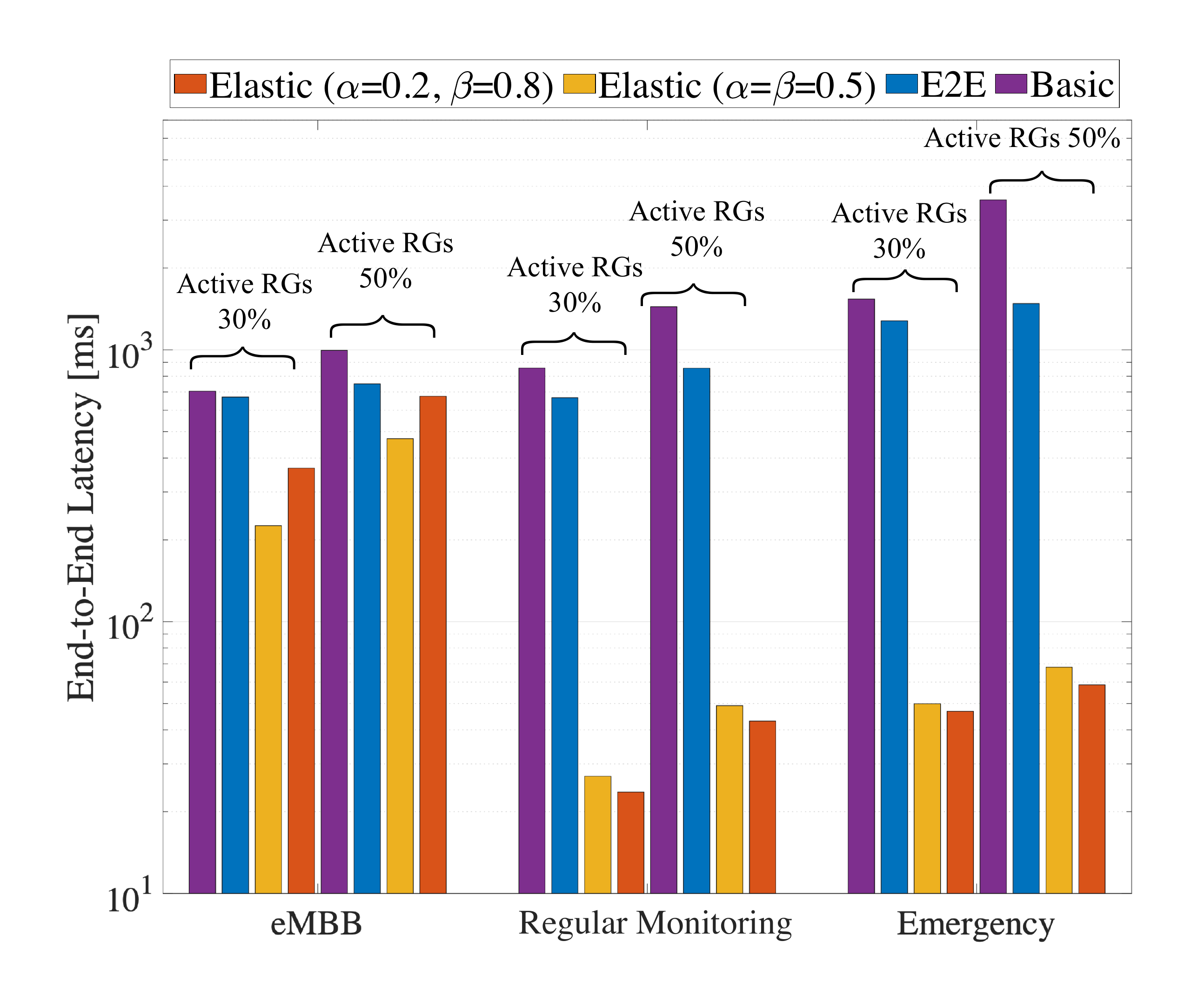}}
 \caption{Average end-to-end latency when 30\% of the RGs (low traffic load) and 50\% of the RGs (high traffic load) are active.}
 \label{fig:delay-6}
\end{figure}	

First, we consider the average end-to-end latency (from the in-home device to the gNB), taking into account the delay introduced by the radio interfaces of the two networks for the considered slice types.

Fig. \ref{fig:delay-6} shows the average end-to-end latency for the various slice types, the different scheduling techniques, and two traffic loads (30\% and 50\% active \glspl{rg}). We observe that both the basic and the \gls{e2e} scheduling yield a higher latency  when the emergency slices are active, since these scheduling approaches do not distinguish among the different traffic types and the penalty incurred by having   higher loads is shared equally among all applications. Moreover, as \gls{embb} traffic is higher than healthcare traffic, the proportional fair scheduling penalizes the regular monitoring and emergency slices. This effect is just slightly mitigated in the \gls{e2e} approach that uses \gls{m-lwdf}  scheduling, being more sensible to the packet latency. When  elastic scheduling in considered, instead, regular monitoring and emergency traffics are served with a much lower latency, as required by their specifics. In fact, we guarantee a smaller delay than with other strategies, by dropping expired packets. At the same time, we also note that the elastic scheduling slightly reduces the latency of the \gls{embb} slice type.  
Moreover, by adjusting the values of $\alpha$ and $\beta$, we can further control the priority of the healthcare slices with respect to the \gls{embb} slice.
When considering different loads (30\% and 50\% active \glspl{rg}), we note that latency grows with the load for all slice types, when the basic and the \gls{e2e} schemes are used. Instead, when the elastic solution is adopted, the latency changes only slightly for the emergency slice type, thus ensuring the required \gls{qos} anyway. This confirms the robustness of our solution to the traffic load (i.e., the percentage of active \glspl{rg}).

\subsection*{Communication Service Availability}

According to 3GPP, the communication service availability is the ratio between the time wherein the service is delivered according to an agreed \gls{qos} and the time expected to deliver it. In our scenario, the system is considered unavailable whenever a message is not received within the survival time (the sum of the end-to-end latency and the jitter), which is considered as the maximum acceptable delay.

\begin{figure}
 \centering
  \includegraphics[width=\linewidth]{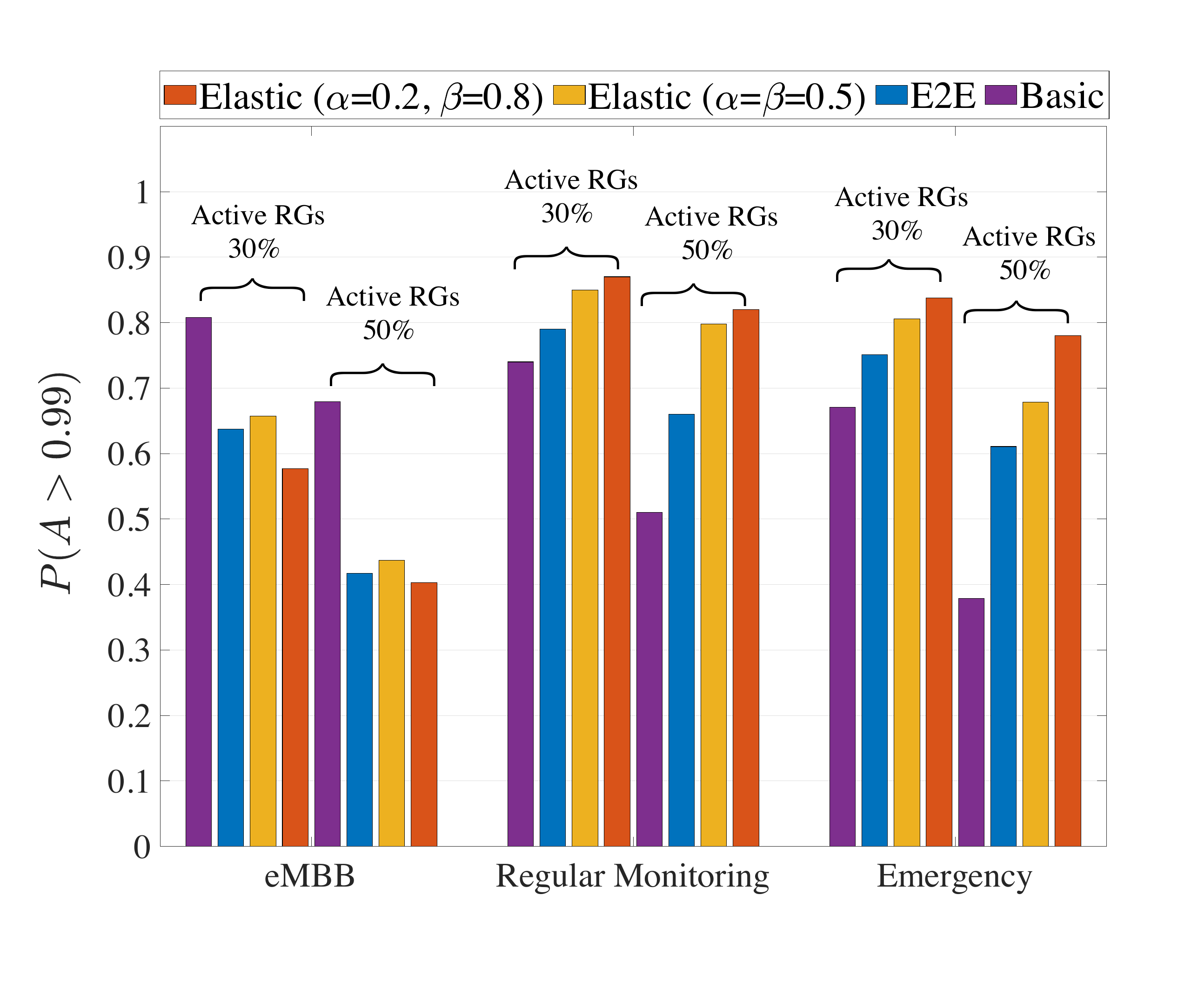}
  \caption{Probability that the communication service availability ($A$) is larger than 0.99 when 30\% and 50\% of the RGs are active (low and high traffic loads, respectively).}  \label{fig:availability}
\end{figure}

Fig.~\ref{fig:availability} shows the probability that the service availability, namely $A$, is larger than 0.99, thus matching the requirements of Table~\ref{tab:scenarios}.  First, we note that the availability dramatically decreases for both regular and emergency slices when the basic solution is used, since the proportional fair scheduling penalizes the healthcare slices under a heavier load (50\% of \glspl{rg}). The \gls{e2e} solution clearly provides higher availability for both healthcare slices, since the wireless networks are sliced and the \gls{m-lwdf} is adopted to schedule the traffic, but it fails to provide decent performance for higher loads (50\% of \glspl{rg}). Our elastic approach, instead, ensures high availability for both healthcare slices in all load conditions (both 30\% and 50\% of active \glspl{rg}), at the cost of a reduced availability of the \gls{embb} slice under a high load (50\% of \glspl{rg}), due to the limited resources of the network. It is important to highlight that the difference, in terms of performance, between the  elastic and the \gls{e2e} strategies, is due to, on one hand, the elastic scaling of resources, and, on the other, the expired packets dropping.

\subsection*{Packets meeting the QoS}

\begin{figure}
  \centering
 \includegraphics[width=\linewidth]{{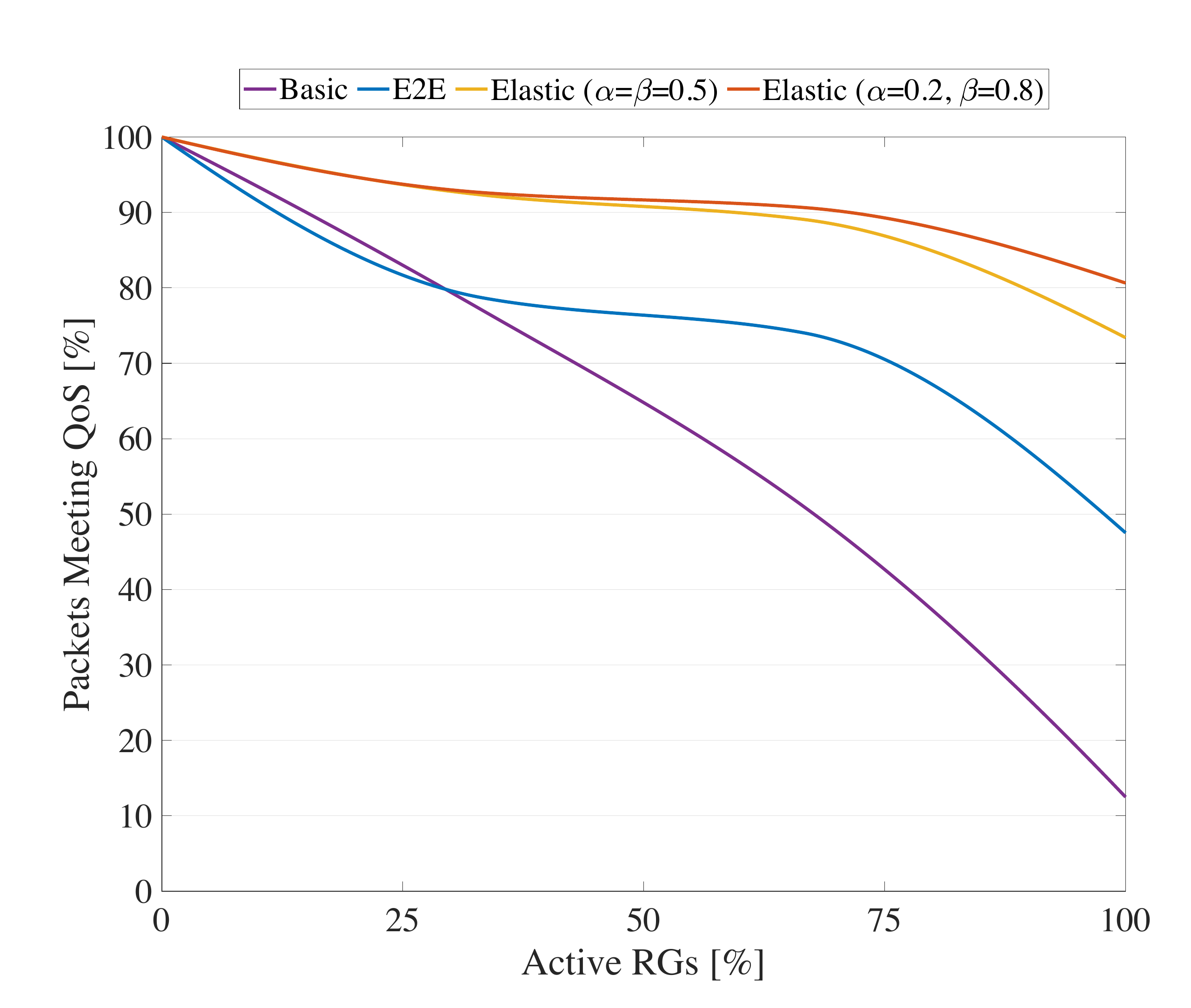}}
  \caption{Percentage of the  \emph{emergency} slice packets meeting the required QoS.}
 \label{fig:qos}
 \end{figure}

To provide a further insight, we consider the percentage of the \emph{emergency} slice packets meeting the required \gls{qos}, as shown in Fig. \ref{fig:qos}.

We note that, when the basic scheme is used, the percentage of packets meeting the \gls{qos} requirements decreases linearly with the percentage of active \glspl{rg}, being thus inadequate to support the healthcare traffic. The \gls{e2e} solution yields a performance improvement, but it still has a linear decay for a small percentage of active \glspl{rg}, with the percentage of packets meeting the \gls{qos} quickly dropping below 80\%. Hence,  also \gls{e2e} is not a  scalable solution. The elastic scheduling technique instead keeps the percentage of   \emph{emergency}  packets meeting the \gls{qos} above 90\% up to 75\% of active \glspl{rg}, with an overall slow decay. Even when all \glspl{rg} are active, the elastic solution ensures that 80\% of emergency packets meet the \gls{qos} requirement, while \gls{e2e} scheme supports less than half of the emergency packets and the basic approach properly serves only 11\% of the packets.

We can then conclude that our solution, based on new slice types and elastic scheduling, is robust to the traffic load, ensuring the required \gls{qos} for healthcare services.

\section*{Conclusions}\label{sec:conclusions}

We have proposed a new communication and computing architecture for pervasive healthcare, with special focus on the indoor monitoring of severe epileptic patients, over a cascaded \gls{wlan}-5G network. The architecture includes several novel elements, including the extension of the slicing concept to \gls{wlan},  the use of new in-network computing capabilities (\gls{gcs}), and the introduction of two new slice types, namely \emph{regular monitoring}  and  \emph{emergency}. Moreover, both a new end-to-end dynamic slicing and resource scheduling across the two networks have been proposed to handle both regular health monitoring and emergency traffic generated by chronic severe epileptic patients living in their smart homes.
%
%
 
%
 
%
Numerical results confirm that our proposed architecture satisfies the diversified \gls{qos} requirements for different slice types and paves the way for future, even more demanding, pervasive healthcare applications.

\section*{Acknowledgments}

All authors are also with CNIT, the National, Inter-University Consortium for Telecommunications. Part of this work was supported by MIUR (Italian Minister for Education) under the initiative {\it Departments of Excellence} (Law 232/2016) and the PRIN project no. 2017NS9FEY entitled "Realtime Control of 5G Wireless Networks: Taming the Complexity of Future Transmission and Computation Challenges".

\bibliographystyle{IEEEtran}
\bibliography{biblio}

\section*{Biographies}
 
{\bf Sergio Martiradonna} (sergio.martiradonna@poliba.it) 
received a Double Degree in Internet Engineering from Politecnico di Bari, Italy, and from Universitè de Nice-Sophia-Antipolis, France, in 2017. He is currently a Ph.D. Student in Electrical and Information Engineering at Politecnico di Bari, Italy. His research interests include 5G, 4G and NB-IoT networks, as well as system-level cellulasimulators.

{\bf Giulia Cisotto} (giulia.cisotto.1@unipd.it) received her Ph.D. in Information Engineering from the University of Padova, Italy, in 2014. Currently, she is Assistant Professor at the Department of Information Engineering of the University of Padova. She served as TCP session chair at the IEEE Healthcom in 2018 and at the IEEE EMBC in 2019. Her current research interests include signal processing and machine learning for electrophysiological signals, wireless body area sensor networks for e-health and Internet-of-Things for healthcare. 
\balance

{\bf Gennaro Boggia} (gennaro.boggia@poliba.it) 
received, with honors, the Dr. Eng. and Ph.D. degrees in electronics engineering, both from the Politecnico di Bari, Bari,
Italy, in July 1997 and March 2001, respectively. Since September
2002, he has been with the Department of Electrical and Information
Engineering, Politecnico di Bari, where he is currently Full
Professor. His research interests include the fields of Wireless Networking, Cellular Communication, IoT, Network Security, Security in Iot, Information Centric Networking, Protocol stacks for industrial applications, Internet measurements, and Network
Performance Evaluation. He is currently an Associate Editor for the IEEE Commun. Mag., the ETT Wiley Journal, and the Springer
Wireless Networks journal.

{\bf Giuseppe Piro} (giuseppe.piro@poliba.it) is an Assistant Professor at Politecnico di Bari, Italy. He has a PhD in electronics engineering from Politecnico di Bari and his main research activities include mobile and wireless networks, network simulation tools, network security, Information Centric Networking, nano communications, and Internet of Things. He is regularly involved as member of the TPC of many prestigious international conferences. Currently, he serves as Associate Editor for Internet Technology Letter (Wiley), Wireless Communications and Mobile Computing (Hindawi), and Sensors (MDPI).

{\bf Lorenzo Vangelista} 
(lorenzo.vangelista@unipd.it)
received the Laurea and Ph.D. degrees in electrical and
telecommunication engineering from the University
of Padova, Padova, Italy, in 1992 and 1995, respectively.
Since October 2006, he has been an Associate Professor of Telecommunication with the Department of Information Engineering, Padova University.
His research interests include signal theory, multicarrier modulation techniques, cellular networks and wireless sensors and actuators networks.

{\bf Stefano Tomasin} (stefano.tomasin@unipd.it) is Associate Professor with the Department of Information Engineering of the University of Padova, Italy. His current research interests include physical layer security and signal processing for wireless communications, with application to the 5th generation of cellular systems. In 2011-2017 he was Editor of the IEEE Transactions of Vehicular Technologies and since 2016 he has been Editor of IEEE Transactions on Signal Processing. Since 2011 he has also been Editor of EURASIP Journal of Wireless Communications and Networking.

\end{document}